\documentclass{article}
\def\R{{\bf R}}
\usepackage{amsmath}
\usepackage{graphicx}

\begin{document}

\title{Multiple steady states in a mathematical model for interactions between 
T cells and macrophages}

\author{Alan D. Rendall\\
Max-Planck-Institut f\"ur Gravitationsphysik\\
Albert-Einstein-Institut\\
Am M\"uhlenberg 1\\
14476 Potsdam, Germany}

\date{}

\maketitle

\begin{abstract}
The aim of this paper is to prove results about the existence and stability
of multiple steady states in a system of ordinary differential equations 
introduced by R. Lev Bar-Or \cite{levbaror} to model the interactions between 
T cells and macrophages. Previous results showed that for certain values of 
the parameters these equations have three stationary solutions, two of which 
are stable. Here it is shown that there are values of the parameters for which 
the number of stationary solutions is at least seven and the number of stable
stationary solutions at least four. This requires approaches different to 
those used in existing work on this subject. In addition, a rather explicit 
characterization is obtained of regions of parameter space for which 
the system has a given number of stationary solutions.
\end{abstract}

\section{Introduction}

Many phenomena in biology and medicine can be modelled by systems of 
ordinary differential equations. Sometimes models are used which have very 
large numbers of unknowns and parameters. Often quantitative information
about these parameters is obtained from high throughput experimental 
techniques. This kind of approach plays a central role in the field of 
systems biology. (For an introduction to systems biology see \cite{klipp}.)
A complementary approach is to try to capture important effects with a
description in terms of dynamical systems with a few unknowns. At the same
time large numbers of parameters may be handled by looking for 
qualitative features of the solutions which are present for large regions
of parameter space. This leads to mathematical problems more accessible to
analytical treatment and less dependent on large-scale simulations. Some 
thoughtful remarks on the relative advantages of these different paths to 
understanding in the applications of mathematics
to biology can be found in \cite{may}. 

One type of qualitative behaviour which is frequently of
interest for biologists is that of multistability, i.e. the existence of
several stable steady state solutions for given values of the parameters.
This means that a biological system can function in more than one way,
depending on its history. In addition to the intrinsic interest of this 
fact it opens up the possibility of manipulating the biological system in 
some way which is favourable for practical applications such as therapies for
certain diseases.

The subject of what follows is a system of four ordinary differential equations
with eighteen parameters introduced in \cite{levbaror} to model certain aspects 
of the immune system. A mathematically rigorous analysis of some featues
of the behaviour of solutions of this system was carried out in 
\cite{rendall10}. In particular it was proved that for some values of the 
parameters there is only one stationary solution while for other parameter 
values there are three stationary solutions, two of which are stable. When 
there is bistability the two stationary solutions can be identified with two 
different states of the immune system which are said to be  Th1-dominated and 
Th2-dominated, respectively. They are distinguished by the concentrations of
certain substances called cytokines which immune cells use to communicate
with each other. In the derivation of the model the cells which produce
and react to the cytokines are T cells and macrophages. The unknowns in
the ODE system are concentrations of cytokines and the cells do not occur
explicitly in the model.

The function of the immune system is to defend the host against harmful
influences such as pathogens and toxins which can cause diseases. (For a
detailed introduction to immunology see \cite{murphy} or \cite{roitt}.) 
Under some circumstances the immune system may malfunction by attacking host
tissues, leading to autoimmune diseases. The choice between Th1- and
Th2-dominated immune responses may have an important influence on the
course of diseases, whether they are due to pathogens or autoimmune in
nature. More information on this can be found in \cite{rendall10} and 
\cite{levbaror}. A survey of work on the development of Th1 and Th2 responses
in an individual organism and approaches to understand this phenomenon using 
mathematical modelling is given in \cite{callard07}.

In what follows it is shown that for suitable values of the parameters
the system of \cite{levbaror} has seven stationary solutions, four of
which are stable. This means for instance that it is possible to have two 
steady states which are Th1-dominated with the degree of dominance being 
different. This property of the system had not previously been observed. 
As noted in \cite{rendall10}, the property of bistability can be
obtained if only T cells are taken into account. On the other hand the 
influence of the macrophages is essential in order to fulfil the assumptions
of the theorems in what follows which assert the existence of more than two 
stable steady states. 

The essential qualitative features of the dynamics exhibited in what follows
can already be found in a much simpler system which arises in a special case.
This model system has only two unknowns and two parameters. The basic system
of \cite{levbaror} and the two-dimensional model system are introduced in 
Sect. \ref{basiceq}. Sect. \ref{stat} contains statements and proofs of 
theorems on multistability for the model system. This allows basic ideas to be 
presented in a relatively simple context. Some of the techniques used extend
in a straighforward way to the analysis of the four-dimensional system. In 
fact they suffice to prove the existence of four stable stationary solutions 
for an open set of the parameter space which is described rather explicitly. 
To prove the existence of more steady states of the four-dimensional system a 
significant refinement of the techniques is necessary. These ideas are 
presented in Sect. \ref{full}. The last section is devoted to further 
discussion of the results of the paper and possible generalizations.

\section{The basic equations}\label{basiceq}

The basic dynamical system studied in this paper is
\begin{equation}\label{basic}
\frac{dx_i}{dt}=-d_ix_i+g(h_i);\ \ \ \ i=1,2,3,4.
\end{equation}
The $d_i$ are positive constants. The function $g$ is given 
by 
\begin{equation}
g(x)=\frac12 (1+\tanh (x-\theta))
\end{equation}
where $\theta$ is a constant. The functions $h_i$ are defined by 
$h_i=\sum_{j}a_{ij}x_j$ for some constants $a_{ij}$. The equations were written 
in this form in \cite{rendall10}. The system was first introduced in 
\cite{levbaror} in a different notation. The relation between the two 
notations is explained in \cite{rendall10}. The coefficients in (\ref{basic}) 
are required to satisfy a number of conditions which will now be listed. The 
formulation of these conditions has been changed slightly from that 
used in \cite{rendall10} so as to make certain calculations more efficient. In 
the formulation used here each $a_{ij}$ is of the form $b_{ij}+c_{ij}$ and
the following conditions hold:
\begin{enumerate}
\item{$(-1)^{i+j}b_{ij}>0$}
\item{$b_{1j}=-b_{2j}$ and $b_{3j}=-b_{4j}$ for all $j$}
\item{$c_{3j}=c_{4j}=0$ for all $j$}
\item{there is a constant $K>0$ such $c_{2j}=Kc_{1j}$ for all $j$.}
\item{$c_{11}\ge 0$ and $c_{13}\ge 0$}
\end{enumerate}
The last of these conditions is a slight weakening compared to \cite{levbaror},
where strict inequality was required. The analysis done in what follows only 
requires the assumptions 1. and 3. Conditions 2., 4. and 5. have been included
here only to make the relation to the original set-up of \cite{levbaror} 
clear.  

The interpretation of these equations is as follows. The quantities $x_1$ and 
$x_3$ are the concentrations of Th1 cytokines produced by T cells and 
macrophages respectively. The quantities $x_2$ and $x_4$ are the 
corresponding concentrations of Th2 cytokines. Thus $z_1=x_1+x_3$ and
$z_2=x_2+x_4$ are the total concentrations of Th1 and Th2 cytokines,
respectively. The signs of the $b_{ij}$ reflect the fact that each type
of cytokine (Th1, Th2) stimulates cells to produce that type of cytokine
and inhibits their production of the other type. The coefficients $c_{ij}$
encode the effect of antigen presentation. The cytokines modify the way in
which macrophages present antigen to T cells and this in turn modifies the
cytokine production of the T cells. The signs of these coefficients are
based on experimental data reviewed in \cite{levbaror}. Setting the 
coefficients $c_{ij}$ to zero corresponds to neglecting the effect of 
antigen presentation. 

A special case of this system which is important in what follows is 
obtained by requiring that $d_i=1$ for all $i$, $\theta=0$, the quantities
$|b_{ij}|$ with $j$ odd are all equal, the quantities $|b_{ij}|$ with $j$ even 
are all equal and the quantities $c_{ij}$ with $i=1,2$ are all equal. Then 
defining $A=b_{11}$, $B=b_{12}$ and $C=c_{11}$ gives a
system which is (21)-(24) of \cite{rendall10}. It implies a closed system 
for the variables $z_1$ and $z_2$ which is (25)-(26) of
\cite{rendall10}. The special case $A=B$ of the last system is given by
\begin{eqnarray}
&&\frac{dz_1}{dt}=-z_1+g((A+C)z_1+(-A+C)z_2)+g(A(z_1-z_2)),\label{basic1}    \\
&&\frac{dz_2}{dt}=-z_2+g((-A+C)z_1+(A+C)z_2)+g(-A(z_1-z_2))\label{basic2}
\end{eqnarray}
and is a useful model system for gaining intuition about the behaviour of
solutions of (\ref{basic}). The constant $A$ is positive while 
$C$ is non-negative. This system is symmetric under interchange of $z_1$ 
and $z_2$. Suppose that $(x_1^*,x_2^*,x_3^*,x_4^*)$ is a stationary solution
of (\ref{basic}) with the restrictions on the parameters $a_{ij}$ leading
to (\ref{basic1})-(\ref{basic2}) and let $z_1^*=x_1^*+x_3^*, z_2^*=x_2^*+x_4^*$. 
Then $(z_1^*,z_2^*)$ is a stationary solution of (\ref{basic1})-(\ref{basic2}) 
and
\begin{eqnarray}
&&x_1^*=g((A+C)z_1^*+(-A+C)z_2^*),\label{xz1}\\
&&x_2^*=g((-A+C)z_1^*+(A+C)z_2^*),\label{xz2}\\
&&x_3^*=g(A(z_1^*-z^*_2)),\label{xz3}\\
&&x_4^*=g(-A(z_1^*-z_2^*)).\label{xz4}
\end{eqnarray}
Conversely, if $(z_1^*,z_2^*)$ is a stationary solution of 
(\ref{basic1})-(\ref{basic2}) 
and $(x_1^*,x_2^*,x_3^*,x_4^*)$ is defined by (\ref{xz1})-(\ref{xz4}) then
a stationary solution of (\ref{basic}) is obtained. The fact that the starting 
solution was stationary assures the consistency conditions that 
$z_1^*=x_1^*+x_3^*$ and $z_2^*=x_2^*+x_4^*$. Thus proving the existence of
stationary solutions of (\ref{xz1})-(\ref{xz4}) also leads to a proof of
the existence of corresponding stationary solutions of (\ref{basic}).

\section{Existence of stationary solutions}\label{stat}

\begin{figure}
\begin{center}{\includegraphics[height=6cm]{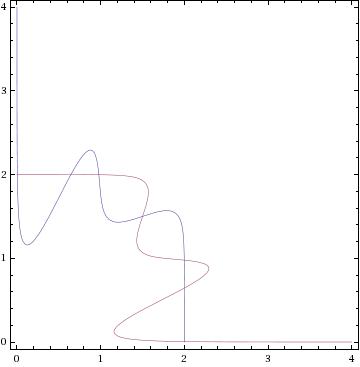}}
\end{center}
\caption{The nullclines of (\ref{basic1})-(\ref{basic2}) in an example
(A=3, C=1.65)}
\label{nullcline}
\end{figure}

In this section results are proved about the existence of stationary
solutions of (\ref{basic1})-(\ref{basic2}). The techniques used to
do so will be generalized in the next section to prove analogous 
results about stationary solutions of (\ref{basic}). A picture which helps 
to provide an intuitive understanding of the results of this section and 
which played an important role in developing the theorems and their proofs is 
given as Fig. \ref{nullcline}. 
The assumptions on the coefficients in 
(\ref{basic}) imply that $C\ge 0$ in the system (\ref{basic1})-(\ref{basic2}) 
derived from it. Nevertheless the possibility that $C$ is negative is allowed 
in this section. The reason for this is that dealing with
negative values of $C$ in (\ref{basic1})-(\ref{basic2}) yields intuition
for the treatment of (\ref{basic}) in the next section where it is allowed 
that some of the coefficients $c_{ij}$ are negative.
For a positive real number $\epsilon$ with $\epsilon<1$ let 
$K_1(\epsilon)=[0,\epsilon]\times[2-\epsilon,2]$,
$K_2(\epsilon)=[2-\epsilon,2]\times[0,\epsilon]$,
$K_3(\epsilon)=[1-\epsilon,1+\epsilon]\times[2-\epsilon,2]$,
$K_4(\epsilon)=[2-\epsilon,2]\times[1-\epsilon,1+\epsilon]$.

\noindent
{\bf Theorem 1} (i) Suppose that $\eta_+<1$ is a positive constant and that 
$\frac{|C|}{A}\le\eta_+$. Then if $A$ is sufficiently large there exists an
$\epsilon>0$ such that there is a unique stationary solution of (\ref{basic1}) 
and (\ref{basic2}) in each of $K_1(\epsilon)$ and $K_2(\epsilon)$. 

\noindent
(ii) Suppose in addition that $\eta_->\frac13$ and $\frac{C}{A}\ge\eta_-$. 
Then for $A$ sufficiently large $\epsilon$ can be chosen such that there is 
a unique stationary solution in each $K_i(\epsilon)$ with $i=1,2,3,4$.

\noindent
{\bf Proof} Because of the symmetry of the system it is enough to prove the
result for $K_1$ and $K_3$. Define a mapping by
\begin{eqnarray}
\phi(z_1,z_2)&&=(g((A+C)z_1+(-A+C)z_2)+g(A(z_1-z_2)), \nonumber\\
&&g((-A+C)z_1+(A+C)z_2)+g(-A(z_1-z_2))).
\end{eqnarray}
Alternatively we can write $\phi(z_1,z_2)=(\phi_1(z_1,z_2),\phi_2(z_1,z_2))$.
Stationary solutions of (\ref{basic1}) and (\ref{basic2}) are in one to one 
correspondence with fixed points of the mapping $\phi$. To prove the theorem 
it is enough to show that for $A$ sufficiently large and a suitable choice of
$\epsilon$ the mapping $\phi$ leaves each $K_i$ invariant and that the 
restriction of $\phi$ to each of these sets is a contraction. Suppose that 
$\epsilon\le\frac14(1-\eta_+)$. Let $A_0$ be a positive constant and assume 
that $A\ge A_0$. The hypotheses of the theorem imply that 
$-A+C\le-(1-\eta_+)A$. Consider first the set $K_1$. For $(z_1,z_2)\in K_1$
\begin{eqnarray}
&&(A+C)z_1+(-A+C)z_2\le (1+\eta_+) A\epsilon-(1-\eta_+) A(2-\epsilon)
\nonumber\\
&&=2A[-(1-\eta_+)+\epsilon]\le -(1-\eta_+) A_0.
\end{eqnarray}
Similarly $A(z_1-z_2)\le A\epsilon-(2-\epsilon)A\le -A_0$. Next the 
arguments of $g$ 
occurring in the second component will be estimated.
\begin{eqnarray}
&&(-A+C)z_1+(A+C)z_2\ge -A(1+\eta_+)\epsilon+A(1-\eta_+)(2-\epsilon)
\nonumber\\
&&=2A\left(1-\eta_+-\epsilon\right) \ge (1-\eta_+)A_0
\end{eqnarray}
and $-A(z_1-z_2)\ge A_0$.
Note that
\begin{equation}
1+\tanh x= \frac{2e^{2x}}{1+e^{2x}},\ \ \ 
1-\tanh x= \frac{2e^{-2x}}{1+e^{-2x}}.
\end{equation}
Hence $g((A+C)z_1+(-A+C)z_2)\le e^{-2(1-\eta_+)A_0}$ and 
$g(A(z_1-z_2))\le e^{-2A_0}$. If $A_0$ is large enough then $\epsilon$ can be 
chosen so that the inequality $e^{-2(1-\eta_+)A_0}\le\frac{\epsilon}2$ is
satisfied. Under these circumstances it follows that $\phi_1(z_1,z_2)$, which 
is evidently positive, is less than $\epsilon$. In the same way 
$g((-A+C)z_1+(A+C)z_2)\ge 1-e^{-2(1-\eta_+)A_0}$ and 
$g(-A(z_1-z_2))\ge 1-e^{-2A_0}$. It follows that with the restrictions on $A_0$
and $\epsilon$ already assumed $\phi_2$, which is 
evidently less than two, is no less than $2-\epsilon$. Hence $K_1$ is 
invariant under $\phi$. To prove that $\phi$ is a contraction it is helpful 
to use the identity $g'(x)=2g(x)(1-g(x))$, which shows that
$g'(x)\le 2\min\{e^{2x},e^{-2x}\}$. It follows by the mean value theorem that 
if $x$ and $y$ both have the same sign and modulus greater than $x_0$ then 
\begin{equation}
|g(x)-g(y)|\le 2e^{-2x_0}|x-y|
\end{equation}
Hence if $(z_1,z_2)$ and $(w_1,w_2)$ both belong to $K_1$ then
\begin{equation}
|\phi_i(z_1,z_2)-\phi_i(w_1,w_2)|
\le 8(1+\eta_+)Ae^{-2(1-\eta_+)A}(\max\{|z_1-z_2|,|w_1-w_2|\})
\end{equation}
for $i=1,2$. The function $xe^{-2(1-\eta_+)x}$  has its unique maximum on the 
interval $[0,\infty)$ at $\frac{1}{2(1-\eta_+)}$. Thus if 
$A_0\ge \frac{1}{2(1-\eta_+)}$ and $A_0e^{-2(1-\eta_+)A_0}<\frac18(1+\eta_+)^{-1}$ 
then the restriction of $\phi$ to $K_1$ is a contraction in the maximum norm.

A similar analysis can be carried out in the case of $K_3$ when the assumptions
of part (ii) of the theorem hold. For $(z_1,z_2)\in K_3$.
\begin{eqnarray}
&&(A+C)z_1+(-A+C)z_2\ge -(1+\epsilon)A+(3-\epsilon)C\nonumber\\
&&\ge A(1+\epsilon)\left[\left(\frac{3-\epsilon}{1+\epsilon}\right)\eta_--1
\right].
\end{eqnarray} 
Suppose that $\epsilon\le\frac{3\eta_--1}{5\eta_-+1}$. Then it follows that 
$(A+C)z_1+(-A+C)z_2\ge\frac{3\eta_--1}{2} A_0$. 
On this region $A(z_1-z_2)\le -\frac12 A_0$.
\begin{equation}
(-A+C)z_1+(A+C)z_2\ge -A(1+\epsilon)+A(2-\epsilon)\ge\frac12 A_0
\end{equation}
and $-A(z_1-z_2)\ge\frac12 A_0$. If it is assumed that 
$e^{-A_0}<\frac{\epsilon}{2}$ and $e^{-(3\eta_--1)A_0}<\frac{\epsilon}{2}$ these
inequalities imply that $K_3$ is invariant. That the restriction of $\phi$ to
$K_3$ is a contraction follows as in the case of $K_1$ under the assumptions
that $A_0\ge\frac{1}{3\eta_--1}$ and 
$A_0e^{-(3\eta_--1)A_0}<\frac18(1+\eta_+)^{-1}$. This completes the proof of 
the theorem.

The symmetry of the system implies that the diagonal $z_1=z_2$ is an invariant
submanifold. Stationary solutions on the diagonal are in one to one 
correspondence with points where $\frac{d}{dt}(z_1+z_2)=0$. When $z_1=z_2$ 
\begin{equation}
\frac{d}{dt}(z_1+z_2)=1-(z_1+z_2)+2g(C(z_1+z_2)).
\end{equation}
Thus stationary solutions on the diagonal are in one to one correspondence with
zeroes of the function $F(x)=1+2g(Cx)-x$ on the interval $(0,4)$. The facts 
that $F(0)=2$ and $F(4)<-1$ show that $F$  has a zero. The function $x-1$ is
monotone increasing while for $C\le 0$ the function $g(Cx)$ is monotone 
non-increasing. Thus in that case the zero of $F$ is unique. For $C>0$ on the 
other hand, when the graph of $2g(Cx)$ crosses that of $x-1$ its derivative 
must be no greater than one. Moreover in this case the second derivative of 
$2g(Cx)$ is negative and so the derivative of $2g(Cx)$ is less than one for all
greater values of $x$. Thus its graph cannot cross that of $x-1$ again. It
follows that the zero of $F$ is unique in all cases and that there is always 
precisely one stationary solution on the diagonal. 
Hence in the cases where Theorem 1 shows the existence of four stationary 
solutions there are at least five in total. Note that if $C$ is large and 
positive then the stationary point is close to $(\frac32,\frac32)$.

\noindent
{\bf Theorem 2} Under the hypotheses of Theorem 1 the system 
(\ref{basic1})-(\ref{basic2}) has at least seven stationary solutions.

\noindent
{\bf Proof} By Theorem 1 and the discussion following it there exist at least 
five stationary solutions. Using
the symmetry of the system it is enough to show that there exists at least
one other stationary solution. Consider the nullclines of the system
(\ref{basic1})-(\ref{basic2}), i.e. the zero sets of the right hand sides of 
the equations. On the first nullcline if $z_1$ is fixed the monotonicity of
$g$ implies that there is at most one possible value for $z_2$. Combining this
with the implicit function theorem shows that the nullcline is the graph
of a function $\gamma_2$ of $z_1$ defined on some open subset. On any closed 
subinterval of this subset the value of $z_2$ is bounded. It follows that in 
fact $\gamma_2$ is defined on the whole interval $(0,2)$. In a similar way
it can be shown that the second nullcline can be written in the form
$z_1=\gamma_1(z_2)$ for a function $\gamma_1$ defined on $(0,2)$. The part 
of the curve $z_1=\gamma_1(z_2)$ where $z_1\le 1+\epsilon$ is contained in the 
strip defined by $2-2e^{-2A_0}\le z_2<2$. This can be seen as follows. From  
estimates very similar to those derived in the proof of Theorem 1 it can be 
concluded that for any fixed value of $z_1$ no greater that $1+\epsilon$ the 
mapping $z_2\mapsto\phi_2 (z_1,z_2)$ defined on the interval $[2-\epsilon,2]$ is
a contraction. The unique fixed point of this mapping lies on the curve
$z_1=\gamma_1(z_2)$ and satisfies the estimate claimed. Let $z_{1,i}^*$ and 
$z_{2,i}^*$ be the $z_1$ and $z_2$ coordinates 
of the stationary solution in $K_i$, respectively. To prove the theorem it is 
enough to show that in the interval $(z_{1,1}^*,z_{1,3}^*)$ there is a value of 
$z_1$ for which $\gamma_2$ is less than $2-2e^{-2A_0}$ and a value of $z_1$ 
for which $\gamma_2$ is greater than $2$. For this it is helpful
to compute the derivative of $\gamma_2$.
\begin{eqnarray}
&&1=[(-A+C)g'((A+C)z_1+(-A+C)z_2)-Ag'(A(z_1-z_2))]\frac{d\gamma_2}{dz_1}
\nonumber\\
&&+[(A+C)g'((A+C)z_1+(-A+C)z_2)+Ag'(A(z_1-z_2))]
\end{eqnarray} 
The second expression in square brackets is small and, in particular, 
less than one half. Taking it onto the left hand side shows that 
$\frac{d\gamma_2}{dz_1}$ is negative and that
\begin{equation}
-\frac{d\gamma_2}{dz_1}\ge
\frac12 [(A-C)g'((A+C)z_1+(-A+C)z_2)+Ag'(A(z_1-z_2))]^{-1}
\end{equation}
It can be concluded that
\begin{equation}
\left |\frac{d\gamma_2}{dz_1}\right|\ge\frac18 A_0^{-1}e^{2(1-\eta_+)A_0}
\end{equation}
in $K_1$ and $K_3$. This guarantees the existence of values of $z_1$ with the 
desired properties and completes the proof.

When the linearization of the system about a stationary point in one of the
$K_i, 1\le i\le 4,$ is computed the contributions of the nonlinear terms are 
extremely small. As a consequence the linearization about a point of this type 
is very close to minus the identity. It follows that the stationary point is a 
hyperbolic sink.

\section{Stationary solutions of the full system}\label{full}

In this section some of the techniques introduced in Sect. \ref{stat} are
applied to obtain information about stationary solutions of (\ref{basic})
which covers an open set of parameters for this system whose definition
is based on explicit 
inequalities. The system of equations defining stationary solutions can
be simplified by replacing the variables $x_i$ by $d_ix_i$.  This means 
that when proving theorems it can be assumed without loss of generality 
that $d_i=1$ for all $i$. Corresponding results for general values of
$d_i$ can then be obtained by transforming the stationary solutions back
to the original variables. First Theorem 1 will be generalized. Note that 
if a stationary solution of (\ref{basic1})-(\ref{basic2}) is close to $(0,2)$, 
$(2,0)$, $(1,2)$ or $(2,1)$ then the corresponding stationary solution of 
(\ref{basic}) defined in Sect. \ref{basiceq} is close to $(0,1,0,1))$, 
$(1,0,1,0)$, $(1,1,0,1)$ or $(1,1,1,0)$ respectively.  For a positive real 
number $\epsilon$ with $\epsilon<1$ define
\begin{eqnarray}
&&\tilde K_1(\epsilon)=[0,\epsilon]\times[1-\epsilon,1]
\times[0,\epsilon]\times[1-\epsilon,1],\nonumber\\
&&\tilde K_2(\epsilon)=[1-\epsilon,1]
\times[0,\epsilon]\times[1-\epsilon,1]
\times[0,\epsilon],
\nonumber\\
&&\tilde K_3(\epsilon)=[1-\epsilon,1]
\times[1-\epsilon,1]\times[0,\epsilon]
\times[1-\epsilon,1],\nonumber\\
&&\tilde K_4(\epsilon)=[1-\epsilon,1]
\times[1-\epsilon,1]
\times[1-\epsilon,1]\times[0,\epsilon].\nonumber
\end{eqnarray}

\noindent
{\bf Theorem 3} (i) Suppose that $d_i=1$ for each $i$, that $\eta_+<1$ and 
$\Delta$ are positive constants, that 
$|c_{ij}|\le \eta_+|b_{ij}|$ for all $i$ and $j$ and that 
$\max_{i,j}\{|b_{ij}|\}\le\Delta\min_{i,j}\{|b_{ij}|\}$. Then if 
$\min_{i,j}\{|b_{ij}|\}$ is sufficiently large there exists an $\epsilon>0$
such that there is a unique stationary solution of (\ref{basic}) in 
each of $\tilde K_1(\epsilon)$ and $\tilde K_2(\epsilon)$. 

\noindent
(ii) Suppose in addition that $\Delta-1$ is sufficiently small, 
$\eta_->\frac13$, $c_{ij}\ge \eta_-\max_{k,l}\{|b_{kl}|\}$ for $i=1,2$ and all
$j$. Then for $\min_{i,j}\{|b_{ij}|\}$ sufficiently large there exists an
$\epsilon$ such that there is a unique stationary solution in each  
$\tilde K_i(\epsilon)$ with $i=1,2,3,4$.

\noindent
{\bf Proof} The proof uses the same strategy as that of Theorem 1. Define a 
mapping $\tilde\phi:\R^4\to\R^4$ by 
\begin{equation}
\tilde\phi_i(x_1,x_2,x_3,x_4)=g\left(\sum_j a_{ij}x_j\right)
\end{equation}
Stationary solutions of (\ref{basic}) are in one to one correspondence with 
fixed points of $\tilde\phi$. To prove the theorem it suffices to show that 
each set $\tilde K_i$ is invariant under $\tilde\phi$ and that the restriction 
of $\tilde\phi$ to each $\tilde K_i$ is a contraction. In comparison with
the proof of Theorem 1 the estimates here give less explicit information about
the conditions in the statement of the theorem. 
Let $A_0$ be a positive constant and assume that $|b_{ij}|\ge A_0$ for 
all $i$ and $j$. Consider first the set $\tilde K_1$. Then 
\begin{eqnarray}
&&\sum_{j}a_{1j}x_j\le -2(1-\eta_+-M\epsilon))A_0,\\
&&\sum_{j}a_{2j}x_j\ge 2(1-\eta_+-M\epsilon))A_0,\\
&&\sum_{j}a_{3j}x_j\le -2(1-M\epsilon))A_0,\\
&&\sum_{j}a_{4j}x_j\ge 2(1-M\epsilon))A_0
\end{eqnarray}
for a constant $M$ depending only on $\Delta$, provided $\epsilon$ is 
sufficiently small. These inequalities imply that $\tilde K_1$ is invariant 
under $\tilde\phi$ if 
\begin{equation}
e^{2|\theta|}e^{-4(1-\eta_+-M\epsilon)A_0}\le\epsilon
\end{equation} 
and this is the case if $A_0$ is sufficiently large and $\epsilon$ is chosen
appropriately. It can be shown that the restriction of
$\tilde\phi$ to $\tilde K_1$ is a contraction using a straightforward
generalization of the argument used in the proof of Theorem 1. Up to a 
numerical factor the contraction constant is
\begin{equation} 
e^{2|\theta|}\Delta \min_{i,j}\{|b_{ij}|\}e^{-4(1-\eta_+-M\epsilon)
\min_{i,j}\{|b_{ij}|\}}
\end{equation}
and this can be made as small as desired by 
choosing $A_0$ large enough. The same arguments can be used to prove that
$\tilde K_2$ is invariant under $\tilde\phi$ and that the restriction of
$\tilde\phi$ to $\tilde K_2$ is a contraction.  

Next part (ii) of the theorem will be proved. This will be presented for the
case of $\tilde K_3$ - the case of $\tilde K_4$ can be treated in the same way. 
Note that if $\xi=1-\Delta^{-1}$ then assuming that $\Delta-1$ is 
sufficiently small is equivalent to assuming that $\xi$ is sufficiently small. 
Moreover for all $i,j,k,l$ the quantity $||b_{ij}|-|b_{kl}||$ is bounded by 
$\xi\max_{i,j}|b_{ij}|$. In this case
\begin{eqnarray}
&&\sum_{j}a_{ij}x_j\ge (3\eta_--1-\xi-6\epsilon)A_0,\ \ \ i=1,2,\\
&&\sum_{j}a_{3j}x_j\le -(1-2\xi-3\epsilon))A_0,\\
&&\sum_{j}a_{4j}x_j\ge (1-2\xi-3\epsilon))A_0.
\end{eqnarray}
This shows that for $\xi$ and $\epsilon$ sufficiently small the set 
$\tilde K_3$ is invariant under $\tilde\phi$. That the restriction to 
$\tilde K_3$ is a contraction can be shown as in the case of $\tilde K_1$. 

\vskip 10pt\noindent
Note that for the parameter values which reduce (\ref{basic}) to 
(\ref{basic1})-(\ref{basic2}) the hypotheses of Theorem 3 reduce to those of
Theorem 1.

The argument used to obtain the stationary solution on the diagonal for the 
system (\ref{basic1})-(\ref{basic2}) does not generalize to the system
(\ref{basic}) since the latter has no symmetry. Instead the observation
concerning the approximate position of that solution for $A$ large made in
the last section will be used. Define
\begin{eqnarray}
&&\tilde K_5(\epsilon)=[1-\epsilon,1]
\times[1-\epsilon,1]\nonumber\\
&&\times\left[\frac12-\epsilon,
\frac12+\epsilon\right]
\times\left[\frac12-\epsilon,\frac12+\epsilon\right].
\end{eqnarray}
The direct analogue of the arguments used to treat $\tilde K_i$ for 
$1\le i\le 4$ does not seem to work for $\tilde K_5$ which is apparently not 
invariant under $\tilde\phi$. To overcome this $\tilde\phi$ will be 
replaced by another mapping $\psi$. 

\noindent
{\bf Theorem 4} Assume that the hypotheses of Theorem 3 hold. Then if
$\Delta-1$ is sufficiently small and $\min_{i,j}|b_{ij}|$ is sufficiently 
large there exists an $\epsilon>0$ such that there is a stationary 
solution of (\ref{basic}) in $\tilde K_5$.

\noindent
{\bf Proof} Define new variables by $y_1=x_1-1$, $y_2=x_2-1$, 
$y_3=x_3-\frac12$ and $y_4=x_4-\frac12$. Then
\begin{eqnarray}
&&b_{31}x_1+b_{32}x_2+b_{33}x_3+b_{34}x_4\nonumber\\
&&=b_{31}(y_1-y_2)+b_{33}(y_3-y_4)+(b_{31}+b_{32})y_2+(b_{33}+b_{34})y_4\nonumber\\
&&+(b_{31}+b_{32})+\frac12 (b_{33}+b_{34})\nonumber\\
&&b_{41}x_1+b_{42}x_2+b_{43}x_3+b_{44}x_4\nonumber\\
&&=b_{41}(y_1-y_2)+b_{43}(y_3-y_4)+(b_{41}+b_{42})y_2+(b_{43}+b_{44})y_4\nonumber\\
&&+(b_{41}+b_{42})
+\frac12 (b_{43}+b_{44})\nonumber
\end{eqnarray} 
The first two terms in each of these expressions are obstructions
to showing that $\tilde\phi$ maps $\tilde K_5$ into itself. To go further note 
that for any positive real numbers $(x,w)$
\begin{equation}\label{xw}
\left|g(x+w)-\frac12-\frac12 x\right|\le\frac12 (x^2+w).
\end{equation}
This suggests rewriting the equation $x_i=g(h_i)$ for $i=3,4$ as
\begin{equation}\label{taylor}
y_i-\frac12 b_{i1}(y_1-y_2)-\frac12 b_{i3}(y_3-y_4)=s_i
\end{equation}
where
\begin{equation}
s_i=g(h_i)-\frac12-\frac12 b_{i1}(y_1-y_2)-\frac12 b_{i3}(y_3-y_4).
\end{equation}
For $i=1,2$ the equation is rewritten as $y_i=s_i$ where $s_1=g(h_1)-1$. These 
equations are schematically of the form $Ny=s$ for some matrix $N$. Conditions 
will now be given which ensure that the matrix $N$ is invertible. For this it 
is useful to write it as a matrix 
$\begin{bmatrix} I & 0 \\ N_{21} & N_{22} \end{bmatrix}$
of two-by-two blocks. If the matrix $N_{22}$ is invertible then $N$ is
also invertible and the inverse of $N$ is given by 
$\begin{bmatrix} I & 0 \\ -N_{22}^{-1}N_{21} & N_{22}^{-1} \end{bmatrix}$
Now consider the matrix $N_{22}$. Its determinant is
\begin{eqnarray}\label{determinant}
&&\left(1-\frac12 b_{33}\right)\left(1+\frac12 b_{43}\right)
+\frac14 b_{33}b_{43}\nonumber\\
&&=\frac14b_{33}^2\left[\left(1-2b_{33}^{-1}\right)
\left(1-2b_{33}^{-1}-\frac{b_{33}+b_{43}}{b_{33}}\right)
-\left(1-\frac{b_{33}+b_{43}}{b_{33}}\right)\right]
\nonumber\\
&&=\frac14b_{33}^2\left[\left(1-2b_{33}^{-1}\right)^2-1
+2\left(\frac{b_{33}+b_{43}}{b^2_{33}}\right)
\right]
\nonumber\\
&&=\frac14b_{33}\left[-4(1-b_{33}^{-1})
+2\left(\frac{b_{33}+b_{43}}{b_{33}}\right)
\right].
\end{eqnarray}
Let $\beta=\min_{i,j}|b_{ij}|$. Then $b_{33}^{-1}\le\beta^{-1}$. It follows in 
particular that if $\beta\ge 2$ then $1-b_{33}^{-1}\ge\frac12$. In addition 
$\left|\frac{b_{33}+b_{43}}{b_{33}}\right|\le\xi$. It follows that 
the square bracket in (\ref{determinant}) is bounded above by 
$-2(1-\xi)$ and from below by $-4(1+\xi)$. In can be concluded that the 
following inequality holds  
\begin{equation}
-\beta\Delta (1+\xi)\le \det N_{22}\le-\frac12\beta(1-\xi).
\end{equation}
In particular, the determinant does not vanish and so $N$ is invertible. 
Moreover the modulus of the inverse of the determinant can be bounded by a 
constant multiple of $\beta^{-1}$. It follows that the norm of $N_{22}^{-1}$, 
can be bounded by a constant and hence that the norm of $N^{-1}$ can be 
bounded by a constant multiple of $\beta$. On the translate of $\tilde B_5$
by $\left(-1,-1,-\frac12,-\frac12\right)$ define a mapping  by 
$\psi(y)=N^{-1}s(y)$. For $i=1,2$
\begin{equation}
|h_i|\ge (2\eta_--2\xi-5\epsilon)(\min_{i,j}|b_{ij}|),\ \ \ i=3,4
\end{equation}
Choose $\xi<\frac12\eta_-$ and $\epsilon<\frac{1}{10}\eta_-$. Then the
first bracket on the right hand side of this equation is greater than
$\frac12 {\eta_-}$. This in turn gives exponential bounds for $s_1$ and $s_2$.
It remains to estimate $s_3$ and $s_4$. Only the argument for $s_2$ will
be given since the corresponding argument for $s_4$ is very similar.
\begin{equation}
|b_{31}(y_1-y_2)+b_{33}(y_3-y_4)|\le 3\beta\Delta\epsilon.
\end{equation}
Using this and (\ref{taylor}) gives the estimate
\begin{equation}
|s_3|\le \frac12 [9\beta^2\Delta^2\epsilon^2+4\beta\xi]
\end{equation}
By choosing $\xi$ and $\beta^2\epsilon$ sufficiently small it can be ensured 
that $s_3$ and $s_4$ are bounded by an arbitarily small multiple of $\epsilon$. 
Now $\epsilon$ can be chosen to satisfy this condition and to ensure at the 
same time that $s_1$ and $s_2$ are bounded by an arbitrarily small multiple of 
$\epsilon$. In this way it can be ensured that $\psi$ leaves $\tilde B_5$
invariant. The conclusion of the theorem follows by applying the Brouwer
fixed point theorem to $\psi$.

\noindent{\bf Remark} This theorem also holds, with the same proof, if the 
hypothesis on $\eta_-$ is weakened to $\eta_->0$.

It has proved possible to extend the method used to obtain Theorem 4 to 
obtain the existence of analogues of the remaining stationary 
solutions of (\ref{basic1})-(\ref{basic2}). Unfortunately this has only been
achieved for a quite restricted choice of parameters. To understand where
the condition for the parameters comes from, note that if 
\begin{eqnarray}
&&\zeta_1=\frac{-b_{12}-c_{12}-b_{14}-c_{14}}{b_{11}+c_{11}},\\
&&\zeta_2=\frac{-b_{21}-c_{21}-b_{24}-c_{24}}{b_{22}+c_{22}}
\end{eqnarray}
then
\begin{equation}
g(h_1(\zeta_1,1,0,1))=g(h_2(1,\zeta_2,1,0))=\frac12
\end{equation}
The cases treated in the next theorem are those where $\zeta_1$ or $\zeta_2$
is close to $\frac12$. Let
\begin{eqnarray}
&&\tilde K_6(\epsilon)=\left[\frac12-\epsilon,\frac12+\epsilon\right]
\times\left[1-\epsilon,1\right]\times\left[0,\epsilon\right]
\times\left[1-\epsilon,1\right],\nonumber\\
&&\tilde K_7(\epsilon)=\left[1-\epsilon,1\right]
\times\left[\frac12-\epsilon,\frac12+\epsilon\right]
\times\left[1-\epsilon,1\right]\times\left[0,\epsilon\right].\nonumber
\end{eqnarray}

\noindent
{\bf Theorem 5} Assume that the hypotheses of Theorem 4 hold and that
\begin{equation}\label{zeta1}
\left|\frac{c_{11}+2c_{12}+2c_{14}}{\beta}-3\right|\le\xi.
\end{equation}
If $\xi$ is small enough and $\beta$ is sufficiently large then $\epsilon$ can 
be chosen so that there is a stationary solution of (\ref{basic}) in 
$\tilde K_6(\epsilon)$. An analogous statement holds if the inequality
(\ref{zeta1}) is replaced by 
\begin{equation}\label{zeta2}
\left|\frac{c_{21}+2c_{22}+2c_{24}}{\beta}-3\right|\le\xi
\end{equation}
and $\tilde K_6(\epsilon)$ is replaced by $\tilde K_7(\epsilon)$ in the 
conclusion.

\noindent
{\bf Proof} The strategy of the proof follows that of Theorem 4. It is enough 
to treat the case of $\tilde K_6(\epsilon)$ since that of 
$\tilde K_7(\epsilon)$ is very similar. This time define new variables by 
$y_1=x_1-\frac12$, $y_2=x_2-1$, $y_3=x_3$ and $y_4=x_4-1$. 
\begin{eqnarray}
&&(b_{11}+c_{11})x_1+(b_{12}+c_{12})x_2+(b_{13}+c_{13})x_3+(b_{14}+c_{14})x_4
\nonumber\\
&&=(b_{11}+c_{11})y_1+(b_{12}+c_{12})y_2
+(b_{13}+c_{13})y_3+(b_{14}+c_{14})y_4\nonumber\\
&&+\left(\frac12 b_{11}+b_{12}+b_{14}+\frac12 c_{11}+c_{12}+c_{14}\right)
\end{eqnarray}
In this case the equation $x_1=g(h_i)$ will be rewritten as
\begin{equation}
y_i-\frac12(b_{11}+c_{11})y_1-\frac12(b_{12}+c_{12})y_2
-\frac12(b_{13}+c_{13})y_3-\frac12(b_{14}+c_{14})y_4=s_1
\end{equation}
where 
\begin{equation}
s_1=g(h_1)-\frac12-\frac12(b_{11}+c_{11})y_1-\frac12(b_{12}+c_{12})y_2
-\frac12(b_{13}+c_{13})y_3-\frac12(b_{14}+c_{14})y_4
\end{equation}
For $i=2,3,4$ the equation $x_i=g(h_i)$ is written as $y_2=g(h_2)-1$, 
$y_3=g(y_3)$ and $y_4=g(h_4)-1$. So as in the proof of Theorem 4 the 
condition for stationary solutions has been rewritten in the form $Ny=s$ for 
a matrix $N$. In this case $N$ is of the form
$\begin{bmatrix} N_{11} & N_{12} \\ 0 & I \end{bmatrix}$
where $N_{11}$ is scalar with the only entry $1-\frac12 a_{11}$, $N_{12}$ is 
a row matrix with entries $-\frac12 a_{1i}$, $i=2,3,4$, and the identity matrix 
is three by three. $1-\frac12 a_{11}\le 1-\frac12\beta$.
If $\beta>3$ the matrix is invertible. The inverse of $N_{11}$ can be bounded
by a constant multiple of $\beta^{-1}$. The inverse matrix $N^{-1}$ is easily 
computed. Its norm can be bounded by a constant multiple of $\beta$. On the 
translate of $\tilde B_6$ by $\left(-\frac12,-1,0,-1\right)$ define a mapping  
by $\psi(y)=N^{-1}s(y)$. It will now be shown that $\psi$ maps its domain into 
itself. The estimates
\begin{eqnarray}
h_2&&\ge (1-\xi-3\epsilon)\beta,\\
h_3&&\le -(1-\xi-3\epsilon)\beta,\\
h_4&&\ge (1-\xi-3\epsilon)\beta
\end{eqnarray}
suffice to take care of the last three components $s_i$. It remains to treat 
the first component. Once this has been done the estimate for $N^{-1}$ which 
is already available completes the proof. The quantity $s_1$ can be estimated
using the inequality (\ref{xw}).
\begin{equation}
|(b_{11}+c_{11})y_1+(b_{12}+c_{12})y_2+(b_{13}+c_{13})y_3+(b_{14}+c_{14})y_1|
\le 8\beta\Delta\epsilon.
\end{equation}
It follows that
\begin{equation}
|s_1|\le 32\beta^2\Delta^2\alpha^2+\left(\frac52\Delta+1\right)\beta\xi
\end{equation}
This estimate leads to the desired conclusion as in the proof of Theorem 4.

For the special choices of the parameters leading to the system
(\ref{basic1})-(\ref{basic2}) the quantities $\zeta_1$ and $\zeta_2$ reduce to
\begin{equation}
\zeta_1=\zeta_2=2\left(\frac{1-\frac{C}{A}}{1+\frac{C}{A}}\right).
\end{equation}
In this case, as $\frac{C}{A}$ varies from one to $\frac13$, the $\zeta_i$ 
vary from zero to one. Each of the extra conditions (\ref{zeta1}) and 
(\ref{zeta2}) on the parameters reduces to $\frac{C}{A}=\frac35$. When Theorem 
5 applies this gives some information about the position of the stationary 
solutions whose existence is guaranteed by Theorem 2.  

\section{Conclusions and outlook}

The dynamical system introduced in \cite{levbaror} to describe the 
interactions between T cells and macrophages has been analysed with respect to
its stationary solutions. For certain open sets of the parameter space it was 
shown that there are at least seven stationary solutions of which at least 
four are stable. The positions of these stationary solutions were described 
approximately. The same features are already found in the model system 
(\ref{basic1})-(\ref{basic2}) with the two parameters $A$ and $C$. There it 
is only necessary to assume that $\frac13<\frac{C}{A}<1$ and that $A$ is
sufficently large to get these conclusions. Under weaker assumptions, which 
reduce to $\frac{|C|}{A}<1$ and $A$ sufficiently large for the model system it 
was shown that there are at least three stationary solutions of which at least 
two are stable. By contrast it was shown in \cite{rendall10} that under a 
suitable smallness assumption on the parameters there is a unique stationary 
solution and that all solutions converge to it at late times. For the model
system a sufficient condition implying the latter behaviour is $2A+|C|<1$.

As already remarked in \cite{rendall10}, for $C<A$ the system 
(\ref{basic1})-(\ref{basic2}) is competitive and hence any solution 
converges to a stationary solution as a consequence of the results of 
\cite{hirsch}. In particular there can be no periodic solutions. This 
conclusion depends on the fact that the system is two-dimensional and so 
cannot be drawn for the system (\ref{basic}), even in the case of parameter 
values for which it is competitive. The results of this paper do not rule 
out the possibility that there are values of the parameters for which the
model system has periodic solutions. They also do not prove that there cannot
be more than seven stationary solutions or more than four stable stationary
solutions for some values of the parameters. Simulations have given no 
indications that either of these phenomena (periodic solutions or extra
stationary solutions) actually occur. 

From the biological point of view one of the most interesting new findings of 
this paper is the coexistence of stationary solutions such as those close to
$(0,1,0,1)$ and $(1,1,1,0)$ with the same values of the parameters. Certain 
small changes of the parameters can move the system from a situation where 
both of these exist to a situation where one of them has disappeared. On a
heuristic level this gives a scenario where a small external influence on
the system leads to a large change in its properties. Note that these two
stationary solutions have very different biological properties. In the second
case the Th1 and Th2 cells are secreting a comparable large amount of 
cytokines. The total cytokine concentration is $50\%$ greater in the second 
case than in the first. The model of \cite{levbaror} is crude in may ways
but their are many models for biological systems which share qualitative
features with this one and the kind of switching phenomenon which has been 
observed here may occur much more widely. Thus it is good to have  
mathematical tools which allow it to be studied.

The Th1/Th2 picture of autoimmune diseases is no longer up to date. Newer
models involve players other than Th1 and Th2 cells. In particular Th17
cells have come to play a very important role \cite{steinman07}.
Mathematical insights obtained for one model can be useful in understanding
other models, even those which remain to be invented. Thus the fact that
the biological content of the model of \cite{levbaror} is not close to the
explanations of the behaviour of the immune system which are presently most
popular does not prevent it being worthwhile to study its mathematical 
properties in detail.  

The mathematical description of the immune system in the model of 
\cite{levbaror} is at the level of the interaction of populations of cells
and what happens in a given cell is handled as a simple black box. 
Another very interesting task is to model the processes within cells which
determine whether they differentiate into types such as Th1 and Th2. These
usually involve the behaviour of transcription factors. Some models of this 
kind are studied in \cite{yates} and \cite{callard07}. Models of the 
population of cells and models of the molecular machinery within cells 
describe systems whose intrinsic nature are very different. Nevertheless 
they may be related mathematically and exploiting this kind of relation 
is a particular strength of mathematics.

\end{document}